\documentclass[journal]{IEEEtran}
\usepackage{cite}
\usepackage{amsmath,amssymb,amsfonts,bm,color}
\usepackage[ruled, lined, linesnumbered, commentsnumbered, longend]{algorithm2e}
\usepackage{graphicx}
\SetKwProg{Init}{Initialization: }{}{}
\usepackage{textcomp}
\usepackage{wrapfig}
\usepackage{enumitem}
\usepackage{float}
\usepackage{graphicx}
\usepackage{subfigure}

\usepackage{letltxmacro}
\LetLtxMacro{\origincludegraphics}{\includegraphics}

\renewcommand{\includegraphics}[2][]{\fbox{\origincludegraphics[#1]{#2}}}
\usepackage{booktabs}
\usepackage{multirow}
\usepackage{array}
\usepackage{tabularx}
\usepackage{hhline}
\usepackage{tikz}
\usetikzlibrary{arrows.meta,positioning,fit,calc,backgrounds}
\newcolumntype{L}[1]{>{\raggedright\arraybackslash}p{#1}}
\DeclareMathAlphabet\mathbfcal{OMS}{cmsy}{b}{n}
\begin{document}
	\title{Physics Equivariance for Robust Generalization in Wireless Foundation Model}
    \author{Haoyu Wang, Xigang Gao, Zhi Sun, and Zhaocheng Wang \vspace{-20pt}
		\thanks{Haoyu Wang, Xigang Gao, Zhi Sun, and Zhaocheng Wang are with the Department of Electronic Engineering, Tsinghua University, Beijing 100084, China (e-mail: wanghy22@mails.tsinghua.edu.cn; gxg25@mails.tsinghua.edu.cn; zhisun@ieee.org; zcwang@tsinghua.edu.cn). Haoyu Wang and Xigang Gao contributed equally to this work. }
	\thanks{Corresponding Author: Zhi Sun}
	}
    % \author{Haoyu Wang, Zhi Sun, Shuangfeng Han, Xiaoyun Wang,  Shidong Zhou, and Zhaocheng Wang \vspace{-20pt}
	\maketitle
			\begin{abstract}
            Wireless foundation models (WFMs) have recently emerged as a promising paradigm for learning multiple channel state information (CSI) acquisition tasks. However, unlike natural language tokens governed by statistical co-occurrence, wireless channels are generated by electromagnetic propagation laws, and current WFM training is constrained by limited data scale, narrow distribution coverage dominated by simulations, and a pronounced sim-to-real gap. As a result, simply scaling model parameters and CSI samples does not necessarily yield robust and generalizable models. In this paper, we advocate enabling physics equivariance as a principled and explainable inductive bias for WFMs. Specifically, we focus on a universal propagation property for electromagnetic waves, termed wave equivariance: when the input CSI is modulated along time–frequency–space dimensions, the output channel response should exhibit the corresponding transformation. Empirical studies show that the vanilla-WFM fails to reliably acquire such equivariance even with a large number of model parameters and training samples. To address this, we design the physics-intrinsic WFM (phys-WFM) with wave equivariance, which explicitly aligns model behaviors with an interpretable wave propagation structure. Results demonstrate that the proposed design effectively captures wave equivariance and substantially improves robustness and generalization to unseen environments under distribution shift, offering a physics-grounded and testable route toward explainable wireless foundation models.
            \end{abstract}
	% \vspace{-20pt}
\section{Introduction}
\label{sec:intro}
% Channel acquisition importance (to be placed before Sec. I-A)
Accurate and low-overhead channel acquisition is a cornerstone for reaping the gains of massive multiple-input multiple-output (MIMO) in 6G. In the 6G era, both the antenna array size and the system bandwidth are expected to further increase to boost spectral efficiency and network capacity \cite{ieeenet_vision_Saad_2020}. To fully exploit the resulting spatial multiplexing and wideband gains, the acquisition of channel state information (CSI) with high accuracy under low pilot overhead is critical, making efficient CSI acquisition a key enabler for next-generation wireless systems.

Deep learning is promising to enable accurate and low-overhead channel acquisition in massive MIMO systems. Thanks to their strong nonlinear approximation capability, deep neural networks can effectively model the complex dependencies of wireless channels across the time, frequency, and spatial domains \cite{vtm_ai_zhang_2023}. However, most existing learning-based CSI acquisition methods are trained for a specific environment and a fixed set of system configurations, and are often designed for a single type of channel acquisition task, making it difficult to jointly model and serve multiple tasks. These limitations reduce their adaptability to highly dynamic wireless scenarios and their flexibility across different parameter settings, thus hindering practical deployment.

\subsection{Wireless Foundation Models}
Wireless foundation models have recently emerged as a new paradigm for moving beyond task-specific, environment-specific channel learning. Inspired by the scaling success of foundation models in natural language processing \cite{kaplan2020scaling}, the wireless community is exploring whether a single large model can serve as the backbone that can be adapted to multiple CSI acquisition tasks and configurations. The overarching goal of the wireless foundation model is to replicate the scaling capability observed in large language models (LLMs) that achieve continual performance improvement by increasing model scale and training data, so as to enable unified learning across environments and CSI acquisition tasks. At a high level, most wireless foundation models adopt transformer-based backbones to flexibly model varying-sized CSI tensors under different system settings (e.g., bandwidth, antenna size, pilot patterns). For instance, a transformer-based masked autoencoder (MAE) structure is adopted in WiFo \cite{liu2025wifo} and HeterCSI \cite{zhang2026hetercsi}. For the training objective, the wireless foundation models typically rely on masked modeling-based self-supervised learning to capture structured dependencies of wireless channels across the time--frequency--space domains. In terms of data, pretraining of the wireless foundation model is conducted on large-scale heterogeneous channel datasets to improve robustness and cross-environment modeling capability. For instance, the WiFo model is pretrained on 19 datasets and the HeterCSI is pretrained on 40 datasets under the 3GPP 38.901 specifications \cite{liu2025wifo, zhang2026hetercsi}, where the datasets are heterogeneous and exhibit different scenarios, system configurations, user speed, etc.

Despite this rapid progress, a key challenge is that most existing wireless foundation models remain purely data-driven and inherit the ``scaling-first'' philosophy from LLMs, where increased data and model size are expected to consistently improve performance. However, such scaling easily hits a bottleneck in the wireless channel. On the one hand, the measured channel data is highly expensive, which is difficult to scale as the corpus data for LLM. On the other hand, the simulated data is faced with the sim-to-real gap, which limits the generalizability. 

\subsection{Proposed physics-intrinsic wireless foundation model}
\label{subsec:phys-intrinsic-wfm}

\begin{figure}[t]
		\centering
        \includegraphics[width=0.45\textwidth]{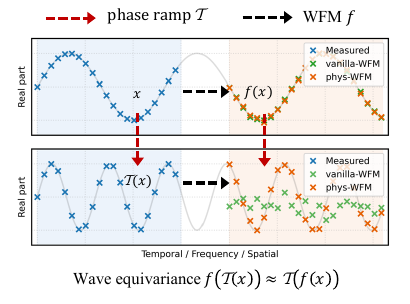}
		% \vspace{-5pt}
		\caption{Illustration of wave equivariance in the proposed phys-WFM.}
        \label{fig: physically_consistent}
		% \label{fig: generalization analysis}
		% \vspace{-15pt}
\end{figure}

To resolve the scaling bottleneck, we propose a physics-intrinsic wireless foundation model (phys-WFM). Our motivation lies in the universality of the electromagnetic (EM) wave propagation mechanism across both simulated and real-world datasets. When the universal mechanism is explicitly embedded, the robustness and generalizability of the wireless foundation model can be significantly enhanced. In the proposed phys-WFM, the central design principle is to use equivariance as a testable signal of physical awareness. Explicitly, for a known physics transformation $\mathcal{T}$, a physically consistent model $f(\cdot)$ should satisfy the transform-in/transform-out relation $f\!\left(\mathcal{T}(\mathbf{x})\right)=\mathcal{T}\!\left(f(\mathbf{x})\right)$. If a model is physically consistent, applying such a transformation to the input representation should lead to a corresponding, known transformation of the predicted channel. This property can be formalized as equivariance and provides an auditable criterion to distinguish physics-consistent learning from data-driven fitting. Equivariance is particularly meaningful for wireless foundation models because it targets universal capability rather than in-distribution accuracy alone. While wireless environments can vary drastically and cause distribution shifts, certain physics-induced structures remain stable across scenarios. Enforcing these structures provides an inductive bias that improves data efficiency and robustness, and also yields interpretability. Explicitly, the model behavior can be explained and verified through its consistency with known physical transformations.

In wireless propagation, many fundamental mechanisms induce predictable input--output behaviors under structured transformations. In this paper, we focus on the wave nature of wireless channels. Then, we identify an environment-stable equivariance termed wave equivariance, which is held with phase-ramp transformation \cite{proakis2007digital} in the temporal, frequency, and spatial domains. As shown in Fig.~\ref{fig: physically_consistent}, under the phase-ramp transformation, the proposed phys-WFM can intrinsically predict the channel when applying the phase-ramp transformation. On the contrary, the vanilla-WFM cannot effectively learn the underlying physical process from the channel data and fails to
accurately model the channel under the transformation. To achieve the goal of equivariance, we develop an asymptotic wave-equivariant module in the proposed phys-WFM. The proposed module provides a theoretical guarantee of the desired equivariance, enabling physically interpretable predictions and improving generalization to unseen environments. 

\subsection{Advantages}
\label{subsec:advs}
The advantages of the proposed phys-WFM can be summarized as follows.
\begin{itemize}
    \item \textbf{Interpretability:} By introducing wave equivariance, the model output under a given physical transformation becomes predictable, making the model behavior easier to understand, verify, and audit.
    \item \textbf{Generalizability:} physics equivariance holds across environments and thus serves as an environment-agnostic inductive bias. Explicitly modeling this relation in a foundation model can significantly improve cross-environment generalization and robustness under practical distribution shifts.
    \item \textbf{Compatibility:} The proposed wave-equivariant module is plug-and-play and compatible with the prevailing multi-dataset, multi-task training paradigm of wireless foundation models.
\end{itemize}

\subsection{Our Contributions}
\label{subsec:contributions}
In this paper, we advocate phys-WFM that embeds environment-stable physical structures into the foundation-model pipeline for CSI learning. Our main contributions are summarized as follows. Firstly, we analyze the rationale that the prevailing scaling-first paradigm can fall short for wireless foundation models, by highlighting both mechanism-level and data-level mismatches between wireless channels and corpora. Secondly, we identify and formalize wave equivariance as an environment-stable, auditable physical criterion (transform-in/transform-out) for assessing and enforcing physical awareness in CSI modeling. Thirdly, we develop the phys-WFM that enables provable wave equivariance while remaining compatible with multi-dataset, multi-task training. Fourthly, the generalization gain of the proposed phys-WFM is explained from the perspective of effective sample coverage. Extensive experiments demonstrate that the proposed module substantially improves physics consistency and enhances robustness and cross-environment, multi-task generalization, especially under distribution shifts and sim-to-real gaps.

\section{Why Scaling Alone Falls Short for Wireless Foundation Models}
The architecture and training paradigm of wireless foundation models are largely inspired by LLMs, specifically the transformer backbones and scaling performance. Nevertheless, it is far from straightforward for wireless foundation models to replicate the scaling success of LLMs, due to fundamental mismatches at both the mechanism level and the data level, as summarized in Table~\ref{tab:comparison}. The following two subsections contrast these mismatches in detail.

\begin{table*}[!t]
\caption{Comparison between LLMs and wireless foundation models}
\label{tab:comparison}
\centering
{\renewcommand{\arraystretch}{1.35}
\begin{tabular}{c|c|c|c}
\hline
\textbf{Level} & \textbf{Aspect} & \textbf{LLMs} & \textbf{Wireless foundation models} \\
\hline
\multirow{2}{*}{\textbf{Mechanism}} 
& Dependency & Token co-occurrence (statistics) & Wave propagation (physics) \\
\cline{2-4}
& Objective & Cross-entropy (likelihood) & Mean square error (regression) \\
\hline
\multirow{4}{*}{\textbf{Data}} 
& Representation & Discrete token sequences & Continuous complex CSI tensors \\
\cline{2-4}
& Scale & Trillion-level tokens & Billion-level CSI points \\
\cline{2-4}
& Source & Diverse, filtered, curated corpora & Limited real data, mostly simulated (e.g., 3GPP) \\
\cline{2-4}
& Coverage & Broad coverage & Limited coverage; sim-to-real gap \\
\hline
\end{tabular}}
\end{table*}

\subsection{Mechanism Level: Physics vs. Statistics}
The mechanism-level distinction implies that directly transplanting the LLM training paradigm may bias wireless foundation models toward overfitting spurious statistical correlations instead of learning physically grounded dependencies that transfer across environments. Specifically:
\begin{itemize}
    \item Dominant dependency: Wireless channel measurements are coupled across time--frequency--space primarily through physical dependencies governed by EM wave propagation, whereas LLMs dominantly rely on statistics-driven token co-occurrence.
    \item Objective: The objective of the wireless foundation model is typically formulated as regression over continuous-valued channels and trained with objectives such as the mean-squared error (MSE) to capture propagation-induced structure, whereas LLMs are trained by maximizing sequence likelihood via the cross-entropy objective \cite{liang2026large}. 
\end{itemize}

% \subsection{Scale Mismatch: Billions vs. Trillions}

\subsection{Data Level: Wireless Channel vs. LLM Corpora}
\label{subsec:dist-mismatch}
At the data level, wireless foundation models and LLMs differ markedly in three aspects, including data representation, data source/scale, and data coverage. These three aspects together shape how far scaling can go in practice.
\begin{itemize}
    \item Representation: Wireless foundation models are typically trained on complex-valued channel measurements such as CSI tensors organized over time--frequency--space and antenna ports. On the contrary, LLMs are trained on corpus, which are represented by discrete token sequences. 
    \item Source and scale: Wireless foundation models are often trained with CSI datasets at only the billion-sample scale, and the data generation heavily relies on channel simulators due to the high cost and limited openness of large-scale real-world channel measurements. For instance, the WiFo model is pretrained on the simulated channel dataset with billion-level CSI data points \cite{liu2025wifo}, which is generated under the specifications of 3GPP 38.901 document \cite{3gpp38901}. LLM pretraining benefits from trillion-scale tokens that are heavily filtered and curated. For instance, the open-source Qwen-3 model is pretrained on a dataset with 36 trillion tokens, covering diverse domains including general knowledge, coding, reasoning, etc. \cite{yang2025qwen3}. 
    \item Coverage: The relative scarcity and limited coverage of wireless channel measurements make data availability a central bottleneck for effective scaling of wireless foundation models. Due to the simplifications in simulated channel model, the pretraining distribution is inherently constrained by the modeling assumptions of the simulator and may not cover the long-tail propagation conditions in practice, leading to a persistent sim-to-real gap and challenging the scalability and generalization of wireless foundation models. On the contrary, modern LLMs are typically pre-trained on corpora at the trillion-token scale, where the raw web/text streams are further processed by sophisticated filtering, deduplication, and dataset curation pipelines to promote diversity. 
\end{itemize}
In summary, the above mechanism- and data-level comparisons suggest that simply scaling model size and training compute may yield limited or unstable gains for robust generalization in wireless foundation models, especially facing the bottlenecks of data availability and coverage. Since wireless channels are governed by universal and structured physical processes, incorporating physics-informed inductive biases into model design and learning becomes a natural and often necessary complement to data-driven scaling.

\begin{figure*}[t]
    \centering
    \includegraphics[width=1\textwidth]{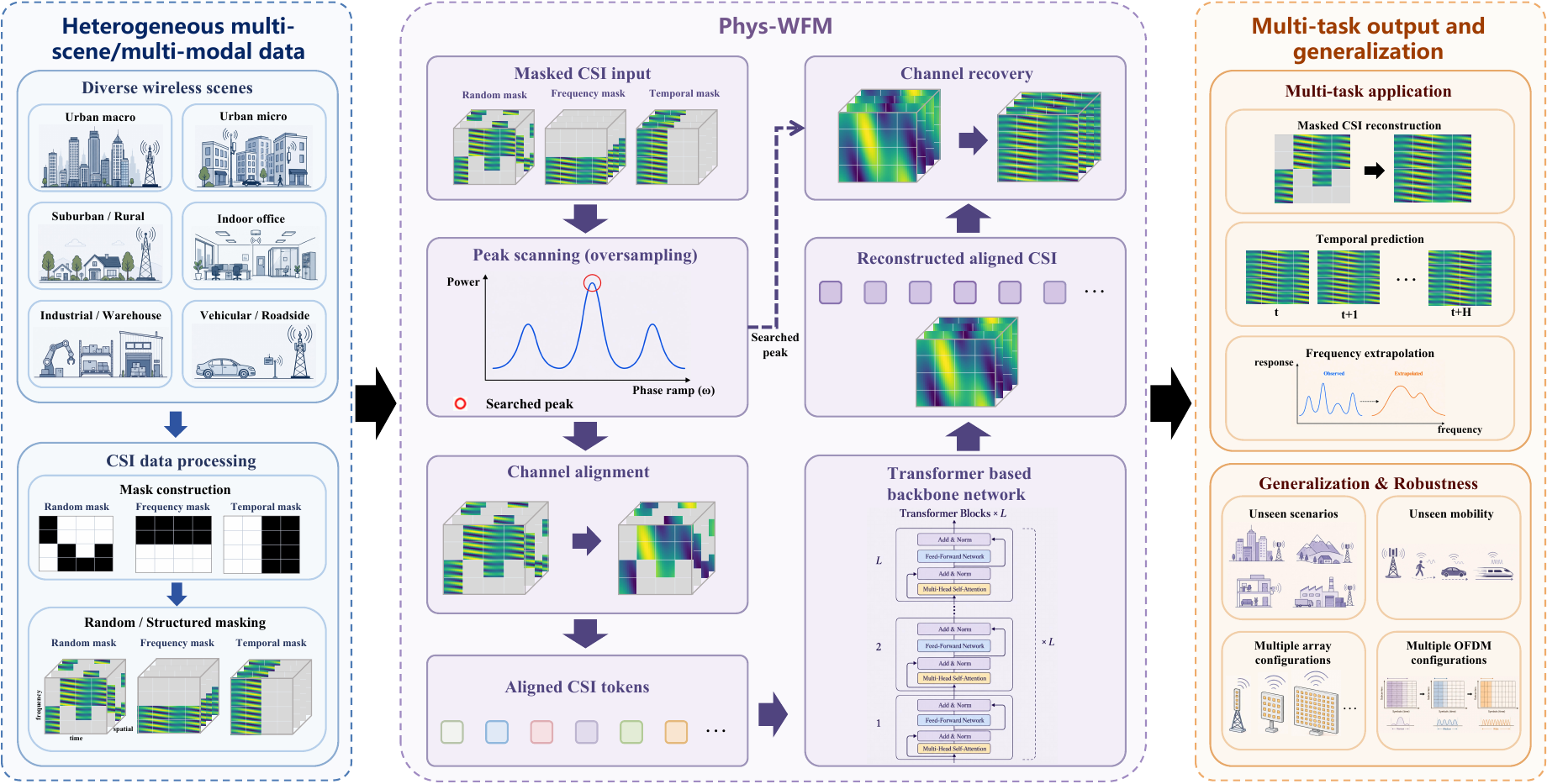}
    \caption{End-to-end structure of the proposed phys-WFM.}
    \label{fig:Asymptotic_Equivariant}
\end{figure*}

\section{Physics-Intrinsic Wireless Foundation Model via Wave Equivariance}
Based on the EM wave propagation, we first introduce the wave equivariance in Sec.~\ref{subsec: wave equivariance}. Next, in Sec.~\ref{subsec: module}, the practical asymptotic wave equivariance is enabled in the proposed phys-WFM. Then, the rationale of generalizability improvement via wave equivariance is presented in Sec.~\ref{subsec: generalization interpretation}.
\subsection{Wave Equivariance}
\label{subsec: wave equivariance}
Wave equivariance is derived from the EM wave propagation law, which is the key to enhancing the generalizability of the proposed phys-WFM. Physically, wave equivariance is grounded in the structured phase progressions induced by multipath propagation parameters across the frequency, spatial, and temporal domains. Specifically, fundamental physical attributes of the wireless channel, namely multipath delay, angle of arrival/departure, and Doppler frequency, dictate these phase progressions within the CSI tensor. Applying a phase-ramp transformation along any of these dimensions preserves the underlying propagation structure while inducing a predictable translation in the delay, angle, or Doppler domain. By embedding this mathematical relationship directly into the model, the wave-equivariant architecture ensures that any phase-ramp transformation of the input CSI systematically translates to a predictably modulated output response, faithfully reflecting physical wave characteristics. 

This underlying propagation structure remains intrinsically stable across diverse deployment environments, serving as a robust, scenario-agnostic physical constraint. While simply scaling model parameters and pretraining datasets often fails to reliably internalize these fundamental wave propagation laws, explicitly enforcing wave equivariance restricts the hypothesis space to physically consistent functions, preventing overfitting to dataset-specific correlations. By embedding this structured inductive bias, the model guarantees predictable output transformation behavior. Conceptually, wave equivariance differs fundamentally from data augmentation in its structural enforcement of physical consistency. While data augmentation merely exposes neural networks to perturbed samples to implicitly approximate consistency, wave equivariance mathematically guarantees this alignment through the model architecture itself. By strictly narrowing the hypothesis space to physically consistent functions, the proposed module provides a robust constraint that directly complements data-driven training paradigms. %\textcolor{red}{removed the discussion of invariance}

\begin{figure*}[t]
    \centering
    \includegraphics[width=1\textwidth]{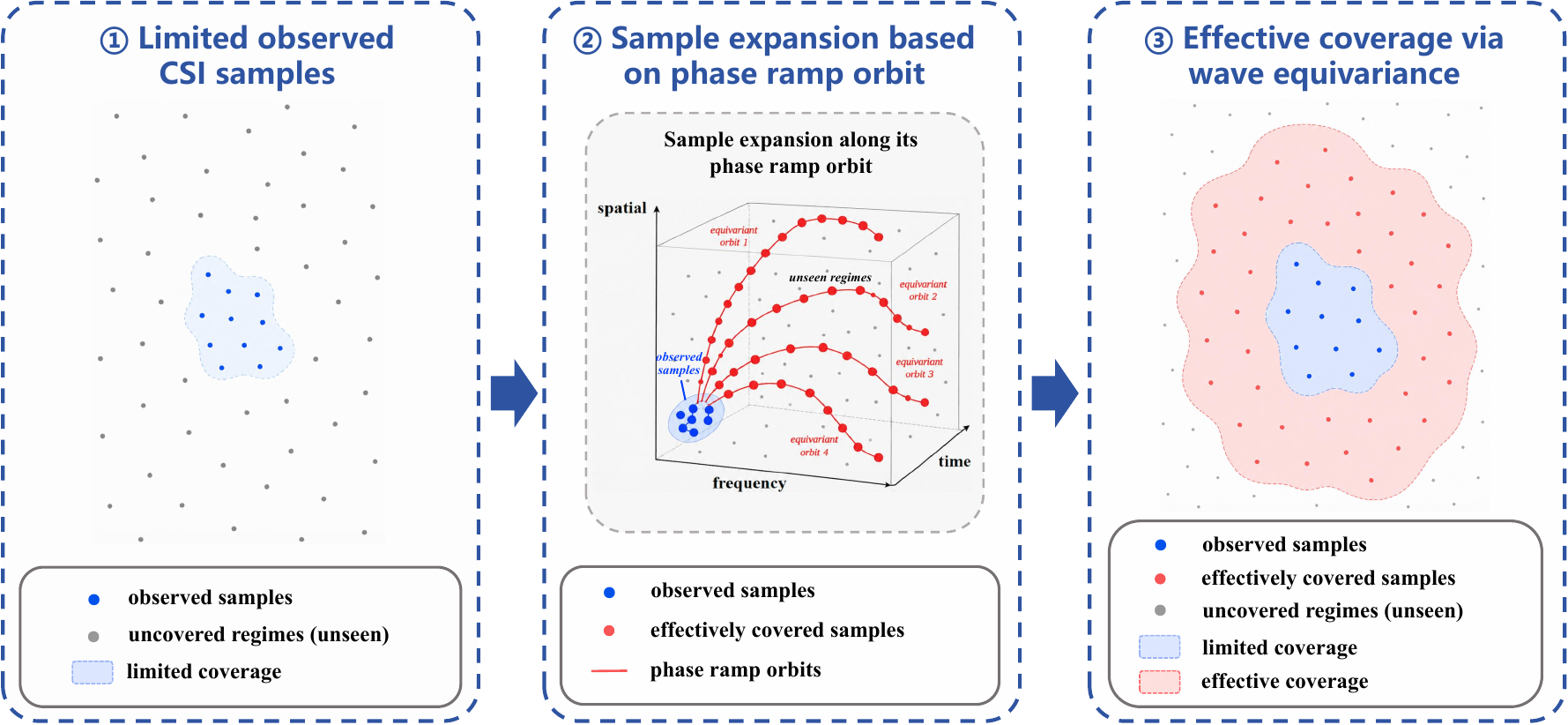}
    \caption{Effective dataset expansion based on wave equivariance, which theoretically improves the generalizability of the wireless foundation model.}
    \label{fig:data_expansion}
\end{figure*}

\subsection{Model Structure}
\label{subsec: module}
The end-to-end structure of the proposed phys-WFM is illustrated in Fig.~\ref{fig:Asymptotic_Equivariant}. Initially, the input CSI undergoes a masking operation, generating a partially observed CSI tensor with missing elements. This masked tensor is then fed into the oversampling-based peak scanning and alignment module to estimate the optimal phase-ramp parameters. By aligning the dominant phase progression of this partially observed channel tensor, the module establishes a standardized coordinate frame to facilitate the backbone processing. Subsequently, the aligned tensor is reconstructed by the transformer-based backbone network \cite{vaswani2017attention} and then mapped back to the original regime by the channel recovery block to fully reconstruct the CSI.

The wave equivariance property in phys-WFM can be achieved asymptotically through a collaborative pipeline of oversampling-based peak search, channel alignment, and channel recovery. Within this framework, the oversampling-based peak search operation enables a precise scanning to capture phase variations in the time, frequency, and spatial domains, which can asymptotically find the ground-truth peak as the oversampling factor increases. Then, the channel alignment module compensates the phase variations in all dimensions based on the scanned peak position. Consequently, even when the input CSI is modulated by arbitrary physical phase ramps, the aligned channel entering the backbone network maintains a relatively stable distribution. This distribution consistency allows the transformer-based backbone to perform stable, high-accuracy completion on the missing elements regardless of physical phase variations. Following backbone completion, the channel recovery block uses the bypassed parameters to apply the inverse transformation, thereby restoring the original channel and realizing end-to-end wave equivariance. Through this explicit physical constraint, the plug-and-play module enhances interpretability, robustness, and cross-scenario generalization.

\subsection{Generalization Gain}
\label{subsec: generalization interpretation}

The wave equivariance property can significantly enlarge the effective coverage of the training channel samples, which is the key to improving model generalization. In wireless foundation models, although scaling model capacity and dataset size remain beneficial, data preparation is constrained by prohibitive measurement costs, leaving pretraining datasets with highly localized coverage, as illustrated in the first part of Fig.~\ref{fig:data_expansion}. To amplify scaling efficiency under such data constraints, wave equivariance exploits physical transformations. Specifically, continuous transformations map a single observed sample into a trajectory of related channel states, forming a sample-level equivariant orbit. By extension, as the raw dataset scales, the collective coverage of the training data expands to the union of these individual orbits. Therefore, the effective physical coverage is multiplied across previously unseen regimes, as depicted in the second part of Fig.~\ref{fig:data_expansion}. This physics equivariance ensures that the network generalizes along these continuous physical orbits, maximizing scaling efficiency while greatly reducing the need for exhaustive real-world measurements across all physical configurations.

This equivariance-driven coverage expansion is particularly crucial for bridging the sim-to-real transfer gap in wireless foundation models. Although simulated, ray-tracing, and physical channels may differ in their marginal distributions, they fundamentally share the same underlying wave propagation properties. Wave equivariance exploits this shared structure to provide an interpretable bridge from observed simulated samples to unseen real-world scenarios.

\section{Experimental Validation}

\begin{figure*}[t]
		\centering		
        \includegraphics[width=1\textwidth]{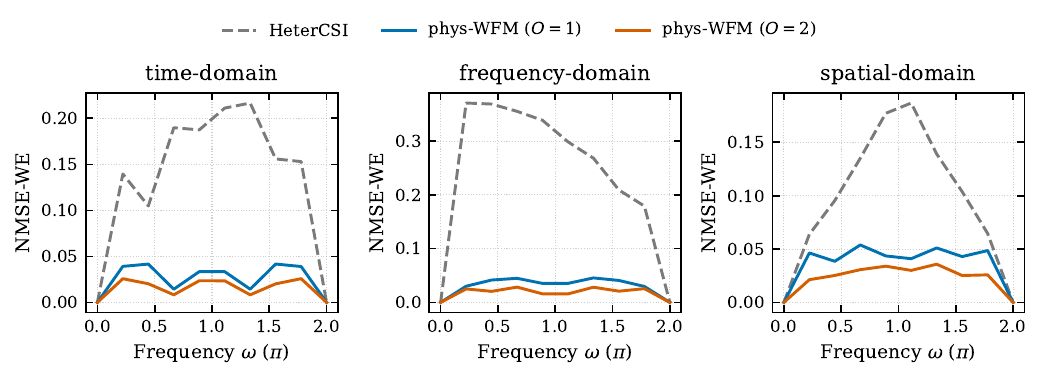}
		% \vspace{-5pt}
		\caption{Wave equivariance comparison over three different domains.}
        \label{fig: eqv}
		% \label{fig: generalization analysis}
		% \vspace{-15pt}
\end{figure*}

% Table generated by Excel2LaTeX from sheet 'Sheet1'
\begin{table*}[t]
  \centering
  \belowrulesep=1pt
  \aboverulesep=1pt
  \caption{Zero-shot generalization comparisons over unseen simulated and real-world channel datasets in dB.}
    \begin{tabular}{c|c|c|c|c|c|c|c|c|c|c|c}
    \toprule
    \multirow{2}[4]{*}{Methods} & \multicolumn{3}{c|}{3GPP} & \multicolumn{3}{c|}{WAIR-D} & \multicolumn{3}{c|}{DeepMIMO} & \multicolumn{2}{c}{RENEW} \\
\cmidrule{2-12}          & Random & Temporal & Frequency & Random & Temporal & Frequency & Random & Temporal & Frequency & Random & Frequency \\
    \midrule
    \multicolumn{12}{c}{Conventional deep learning models} \\
    \midrule
    GRU   & /     & -2.87 & -7.11 & /     & 0.25  & -5.00    & /     & 0.31  & -1.47 & /     & -0.94 \\
    \midrule
    LSTM  & /     & -2.69 & -6.76 & /     & 0.10   & -5.03 & /     & 0.14  & -0.90  & /     & -1.32 \\
    \midrule
    Informer & /     & -3.62 & -6.13 & /     & -0.11 & -6.69 & /     & 0.06 & -3.11 & /     & -0.97 \\
    % \midrule
    % PO-DLE+PA & /     & /     &       & /     & /     &       & /     & /     &       & /     &  \\
    \midrule
    \multicolumn{12}{c}{Wireless foundation models} \\
    \midrule
    WiFo  & -7.28 & -4.71 & -4.75 & -3.79 & -8.99 & -1.38 & -1.97 & -2.38 & -0.49 & -0.47 & 0.59 \\
    \midrule
    HeterCSI & -14.75 & -8.68 & -9.35 & -8.06 & -14.85 & -10.89 & -1.33 & -2.75 & -5.06 & 1.11  & 1.51 \\
    \midrule
    phys-WFM & -15.76 & -9.08 & -9.71 & -17.73 & -19.31 & -14.52 & -13.25 & -14.56 & -12.66 & -3.56 & -3.86 \\
    \bottomrule
    \end{tabular}%
  \label{tab:zeroshot}%
\end{table*}%

\subsection{Experimental Setup}
In the experiments, we adopt MAE as the backbone for phys-WFM. The proposed phys-WFM has 69.9M parameters in total, and is pretrained with a multi-task objective, including random masking reconstruction, time-domain prediction, and frequency-domain extrapolation \cite{liu2025wifo,zhang2026hetercsi}. The pretraining datasets are synthesized based on the 3GPP TR~38.901 channel model \cite{3gpp38901}, including 40 heterogeneous configurations. We compare with representative deep learning baselines, including gated recurrent unit (GRU), long short-term memory (LSTM) \cite{jiang2020deep}, Informer \cite{zhou2021informer}, as well as existing wireless foundation models, including WiFo \cite{liu2025wifo} and HeterCSI \cite{zhang2026hetercsi}.

\subsection{Benchmarking Wave Equivariance}
This subsection quantitatively evaluates the wave equivariance of the proposed phys-WFM against a representative wireless foundation model baseline, HeterCSI, in all three domains, i.e., time, frequency, and space. We quantify the difference between $f(\mathcal{T}_{\omega}(\mathbf{x}))$ and $\mathcal{T}_{\omega}(f(\mathbf{x}))$ with the normalized mean square error (NMSE) metric, which is denoted as NMSE-WE. Here, $\mathcal{T}_\omega$ is the phase-ramp operator with frequency $\omega$. A smaller NMSE-WE indicates better equivariance. Both HeterCSI and our method are trained on exactly the same pretraining dataset. To isolate the equivariance property itself, we then evaluate the above NMSE-WE on the same training set by sweeping the frequency $\omega$ and by testing the three domains separately.

The evaluation results are shown in Fig.~\ref{fig: eqv}. It indicates that the proposed phys-WFM consistently maintains NMSE-WE below $0.05$ with varying $\omega$, demonstrating near-equivariant behavior in time, frequency, and space. Moreover, as the oversampling rate $O$ increases, the NMSE-WE further decreases, which empirically validates the claimed asymptotic property of the wave-equivariant module. In contrast, the baseline HeterCSI exhibits a rapid NMSE-WE increase when $\omega$ approaches $\pi$, indicating that simply scaling up a transformer-based foundation model with masked modeling does not reliably internalize the underlying physical constraints from data alone. This observation supports the conclusion in Sec.~II and verifies the necessity of explicitly enabling physical consistency.

\subsection{Benchmarking Robust Generalization}
We benchmark model generalization in a zero-shot setting, where the trained models are directly evaluated on multiple unseen channel datasets and tasks without any fine-tuning. Specifically, the zero-shot test sets include: (1) unseen 3GPP scenarios/parameter configurations within the standard, (2) the ray-tracing WAIR-D dataset with 100 environments \cite{huangfu2022wair}, (3) the DeepMIMO-cities dataset with 20 environments \cite{alkhateeb2019deepmimo}, and (4) the RENEW measured-channel dataset \cite{du2021massive}. On each dataset, we evaluate three representative channel acquisition tasks with NMSE metric, including random masking-based completion, temporal prediction, and frequency extrapolation.

As shown in Table~\ref{tab:zeroshot}, our proposed phys-WFM achieves the best overall zero-shot accuracy on both simulated and measured datasets, demonstrating a clear sim-to-real advantage. More importantly, it yields the most pronounced gain on the real-world RENEW dataset, indicating that explicitly enforcing wave-propagation equivariance effectively mitigates the sim-to-real gap. Our proposed method also realizes unified multi-task generalization, rather than overfitting to a single acquisition pipeline. Explicitly, the proposed model simultaneously attains the best prediction accuracy on random masking, temporal prediction, and frequency extrapolation across the considered datasets. For instance, the proposed method achieves $-14.52\sim-19.31$ dB on the three tasks in WAIR-D, substantially outperforming the baselines, effectively achieving multi-task and multi-scenario generalizations. This gain is attributed to two complementary factors. On the one hand, benefiting from the scaling advantage, the large amount of training dataset and large model parameter size contribute to the exceptional modeling capability of the model. On the other hand, explicitly modeling wave-propagation equivariance aligns representations across domains and scenarios, thereby significantly enhancing the cross-scenario and cross-task robustness.

The results of the baselines further highlight why physics equivariance is crucial for generalization in wireless foundation modeling. Conventional deep learning baselines typically require training a separate model per task and cannot seamlessly handle heterogeneous CSI dimensions or different parameter configurations, which limits their applicability in open-world settings. Existing wireless foundation models improve flexibility. However, without explicitly modeling the wave-induced equivariance, their performance degrades significantly on unseen distributions, particularly on ray-tracing datasets and the measured RENEW data, which demonstrates the necessity of physics equivariance in wireless foundation models.

\section{Conclusion and Future Directions}

Wireless foundation model is a promising paradigm for scalable channel acquisition in future wireless systems. % However, unlike language tokens whose correlations are largely statistical, CSI exhibits structured dependencies governed by wave propagation, and the data available for WFM pretraining are often limited in both scale and distribution diversity (with a pronounced sim-to-real gap). 
In this paper, we advocate wave equivariance as a principled inductive bias for WFM training and inference, which is a testable property stating that physically meaningful phase-ramp transformations in time/frequency/spatial dimensions should lead to correspondingly transformed channel responses. Then, we propose phys-WFM to achieve the wave equivariance asymptotically, which is compatible with the mainstream transformer-based foundation model backbone. Next, the generalization gain of the proposed phys-WFM is also theoretically explained. Extensive experiments validate that the proposed phys-WFM shows strong zero-shot generalization across unseen heterogeneous channels, thereby providing an interpretable and physically grounded path toward robust CSI modeling.

Although the feasibility and benefits of enabling physics equivariance have been demonstrated, several future directions merit deeper exploration. First, beyond wave equivariance, WFMs may need richer physical structures, e.g., causality-aware temporal evolution, to better cover diverse propagation regimes. Second, the practical effectiveness of equivariance depends on how faithfully physical transformations are preserved under hardware non-idealities and protocol constraints. Thus, deployment-aware design and robustness analysis are important. Finally, a long-term open problem is to build a unified scaling framework where large-scale pretraining and multiple explicit physics priors complement each other, yielding WFMs that are not only accurate but also diagnostically explainable through multiple verifiable physical tests.

\bibliographystyle{IEEEtran}
\bibliography{commag}
    \begin{IEEEbiographynophoto}{Haoyu Wang} [S] (wanghy22@mails.tsinghua.edu.cn)
        received his B.S. degree from Tsinghua University in 2022, where he is currently working toward the Ph.D. degree with the Department of Electronic Engineering, Tsinghua University. His research interests lie in AI for communications.  
    \end{IEEEbiographynophoto} 
    \vspace{-36pt}
    \begin{IEEEbiographynophoto}{Xigang Gao} [S] (gxg25@mails.tsinghua.edu.cn)
        received his B.S. degree from Tsinghua University in 2025, where he is currently pursuing the Ph.D. degree with the Department of Electronic Engineering. His research interests include ISAC and AI for communications.
    \end{IEEEbiographynophoto}
    \vspace{-36pt}
	\begin{IEEEbiographynophoto}{Zhi Sun} [SM] (zhisun@ieee.org)
		 received his Ph.D. degree from Georgia Institute of Technology in 2011. Currently he is a tenured Associate Professor at Tsinghua University, Beijing, China, which he joined in 2021. Prior to that, he was a tenured Associate Professor at University at Buffalo, the State University of New York, USA, which he joined in 2012 as an Assistant Professor. He received the US NSF CAREER Award in 2017. Zhi Sun is the editor for IEEE Transactions on Wireless Communications and Computer Networks (Elsevier). His research interests lie in underground and underwater wireless communications and networking, as well as physical layer security. 
	\end{IEEEbiographynophoto}
    \vspace{-36pt}
    \begin{IEEEbiographynophoto}{Zhaocheng Wang} [F] (wangzc@tsinghua.edu.cn) received his B.S., M.S., and Ph.D. degrees from Tsinghua University, in 1991, 1993, and 1996, respectively. From 1996 to 1997, he was a Post Doctoral Fellow with Nanyang Technological University, Singapore. From 1997 to 2009, he was a Research Engineer/Senior Engineer with OKI Techno Centre Pte. Ltd., Singapore. From 1999 to 2009, he was a Senior Engineer/Principal Engineer with Sony Deutschland GmbH, Germany. Since 2009, he has been a Professor with Department of Electronic Engineering, Tsinghua University. He was a recipient of IEEE Scott Helt Memorial Award, IET Premium Award, IEEE ComSoc Asia-Pacific Outstanding Paper Award, and IEEE ComSoc Leonard G. Abraham Prize.
    \end{IEEEbiographynophoto}
\end{document}